# Characterizing Equitable Access to Grocery Stores During Disasters Using Location-based Data

Authors: Dr. Amir Esmalian[1*], Natalie Coleman[2], Dr. Faxi Yuan[3], Xin Xiao[4], Dr. Ali Mostafavi[5]

[1] PhD, Zachry Department of Civil and Environmental Engineering, Urban Resilience.AI Lab, Texas A&M University, College Station; email: amiresmalian@tamu.edu,
*corresponding author

[2] Urban Resilience.AI Lab, Texas A&M University, College Station; email: ncoleman@tamu.edu

[3] Urban Resilience.AI Lab, Texas A&M University, College Station; email: faxi.yuan@tamu.edu

[4] Urban Resilience.AI Lab, Texas A&M University, College Station; email: xyx56@tamu.edu

[5] Associate Professor, Zachry Department of Civil and Environmental Engineering, Urban Resilience.AI Lab, Texas A&M University, College Station; e-mail: amostafavi@civil.tamu.edu

**Abstract:**
Natural hazards cause disruptions in access to critical facilities, such as grocery stores, impeding residents' ability to prepare for and cope with hardships during the disaster and recovery; however, disrupted access to critical facilities is not equal for all residents of a community. In this study, we examine disparate access to grocery stores in the context of the 2017 Hurricane Harvey in Harris County, Texas. We utilized high-resolution location-based datasets in implementing spatial network analysis and dynamic clustering techniques to uncover the overall disparate access to grocery stores for socially vulnerable populations during different phases of the disaster. Three access indicators are examined using network-centric measures: number of unique stores visited, average trip time to stores, and average distance to stores. These access indicators help us capture three dimensions of access: *redundancy*, *rapidity*, and *proximity*. The findings show the insufficiency of focusing merely on the distributional factors, such as location in a food desert and number of facilities, to capture the disparities in access, especially during the preparation and impact/short-term recovery periods. Furthermore, the characterization of access by considering combinations of access indicators reveals that flooding disproportionally affects socially vulnerable populations. High-income areas have better access during the preparation period as they are able to visit a greater number of stores and commute farther distances to obtain supplies. The conclusions of this study have important implications for urban development (facility distribution), emergency management, and resource allocation by identifying areas most vulnerable to disproportionate access impacts using more equity-focused and data-driven approaches.

**Introduction:**
Natural hazards can disrupt access to healthcare, pharmacies, and grocery stores. Households with easier access to these critical facilities could achieve a higher level of preparation and short-term recovery effort [1,2], and thus, be more capable of withstanding the adverse impact of the disaster [3,4,5]. In particular, better access to grocery stores is critical for access to food and water during and in the aftermath of disasters. Therefore, disrupted access to grocery stores could result in adverse effects on the well-being of residents.

In non-disruptive periods, not all residents enjoy equal levels of access to grocery stores which result in disparate impacts on the well-being of residents in their day-to-day lives. For instance, areas with a dearth of retail establishments selling nutritious food, known as food deserts, are endemic to lower socioeconomic neighborhoods. However, these measures of food inaccessibility are rather limited in examining disparate impacts during disasters when compared to normal conditions [6]. Sub-populations already facing access inequality could be further exacerbated during disasters due to disruptions in road networks [7,8], as well as residents' capabilities and lifestyle patterns [9,10]. Certain vulnerable areas face a greater impact from road closures and damage to the stores, which may disproportionally disrupt the access to the grocery stores[11-14]. To add on, people of a lower socioeconomic status usually have fewer resources and capabilities to find alternatives to food and water to compensate for such disrupted access[15-18].

Indeed, the disaster setting can cultivate a supply-demand imbalance among different sub-populations due to protective actions of residents [3] and disruptions in infrastructure and supply chains [19]. In preparation for an impending disaster, people who choose to shelter-in-place tend to stockpile supplies in anticipation of several days of disruptions with the adverse effect of depleting grocery store inventories [20]. This surge in demand for supplies leads grocery stores to run out of stock and force people to visit multiple stores at farther distances. When faced with local shortages of supplies, households with personal vehicles have the means to travel further to obtain their needs from more distant stores compared to households relying on public transportation. Thus, while the number of stores nearby and the physical distance could provide indications of levels of access during normal times, they are inadequate in examining disparities in accessibility during the preparedness stage. Due to disruptions in road network [21] and power outages, as well as supply chain impacts, the time span for returning to pre-event conditions can be a few weeks to more than a year. The extended period of disruption can significantly impact sub-populations who are unable to access facilities due to increasing trip distance and duration.

The majority of the current literature lacks the nuanced approach to accurately measure accessibility to grocery stores given the unique nature of the disaster context and through physical and social factors. Previous literature defines access based on the number of available stores [6,22-26] or a combination of number of stores and distance from a specified area[21, 27,28]. However, these measures are primarily based on physical distance and only partially characterize access to facilities. Approaches focused on physical distance, as done in the majority of current studies, is not a reliable measure of accessibility during disasters as disruption influences the availability of stores and duration of trips. To begin, less distance from a household to a grocery store may imply better accessibility since the household does not have to travel as far. However, examining distance as an isolated variable limits the holistic perspective needed to understand equitable accessibility. For instance, households living in more affected areas may have a higher number of closed stores due to storm damage which decreases the availability of stores. Flooding events also exacerbate traffic congestion which significantly increases the total trip duration. Although a household is normally a set distance from certain grocery stores, traffic congestion from the closed roads or even construction repairs in damaged roads could significantly increase the travel time. Though these households may have similar distances to grocery stores, they may still spend more time reaching a grocery store to meet their needs. Failure to account for these multiple dimensions of access limits the characterization of access to facilities both during normal times as well as during disasters[6,23,24]. Thus, the research channels the unique challenges brought on by disruption to properly capture the multiple dimensions of access. In contrast to previous approaches, the research will use a combination of access indicators to better understand characteristics of access.

A more nuanced characterization of access based on measures related to population-facility network interactions reveals a more accurate picture of access. Therefore, there is a multitude of physical and social factors in the disaster context that impact the equitable access to grocery stores. The research will specifically focus on the inaccessibility of grocery stores through multiple access indicators while accounting for the specific disproportionate impact on socially vulnerable populations. While the extant literature recognizes importance of access to facilities for community resilience in disasters[29-31], there is limited empirical and observational insights to inform about the impacts of disasters on access to grocery stores[5,32]. In order to address the knowledge gap of using empirical and observational data to capture accessibility, the research presents an innovate method which harnesses and analyses location-based data to examine the equitable access to grocery stores during different phases of the disaster.

In this study, we constructed and analyzed the population-facility network for examining access disparities to grocery stores in the context of Hurricane Harvey in 2017. In reviewing literature of accessibility to critical facilities and functionality of critical systems, we characterized access to facilities based on three distinct but complementary dimensions: (1) *redundancy* in access; (2) *rapidity* of access; and (3) *proximity* of access. Redundancy is based on the number of unique stores visited as in the number of available stores open to consumers. This greater availability of critical facilities can lead to increased potential accessibility to important resources and needed services [22,23,26]. Rapidity of access is based on the duration of trips to stores which accounts for traffic delays and road disruption. Quicker trips can represent the functionality and restoration of the system [33-35] Proximity of access is based on distance to stores visited. Shorter distances could mean that consumers do not have to travel as far to reach important resources [27,28].

Accordingly, this research aims to answer the following research questions: (1) What are the characteristics of access to grocery stores for different sub-populations at different stages of disasters? (2) What is the extent of disparity in access among different sociodemographic groups in different disaster phases? (3) What are the contributing factors to unequal access to grocery stores, and to what extent are facility distribution inequities, such as food deserts, indicative of access disparities? To address these research questions, we defined three distinct indicators for examining access and quantify these indicators based on location-based data related to people's visits to grocery stores in the context of Harris County, Texas.

**Materials and methods:**
We examined disparities in access to grocery stores based on the structure and attributes of population-facility networks. The three steps of this methodology (Fig. 1) were: (1) specifying distinct indicators for examining different dimensions of access, (2) analyzing variations in the access indicators among different sub-populations in three phases of the disaster, and (3) evaluating the factors influencing access disparity. First, we utilized node-level and link-level access metrics to capture different aspects of access in population-facility networks: (1) unweighted degree based on the number of visited grocery stores, (2) weighted degree based on the travel time, and (3) weighted degree based on trip distance. Second, we evaluated fluctuations in patterns of each measure across different spatial areas to find variations in access at different stages of the disaster. Third, we specified disparities in access to grocery stores in the face of a disaster and evaluated factors contributing to such disparities.

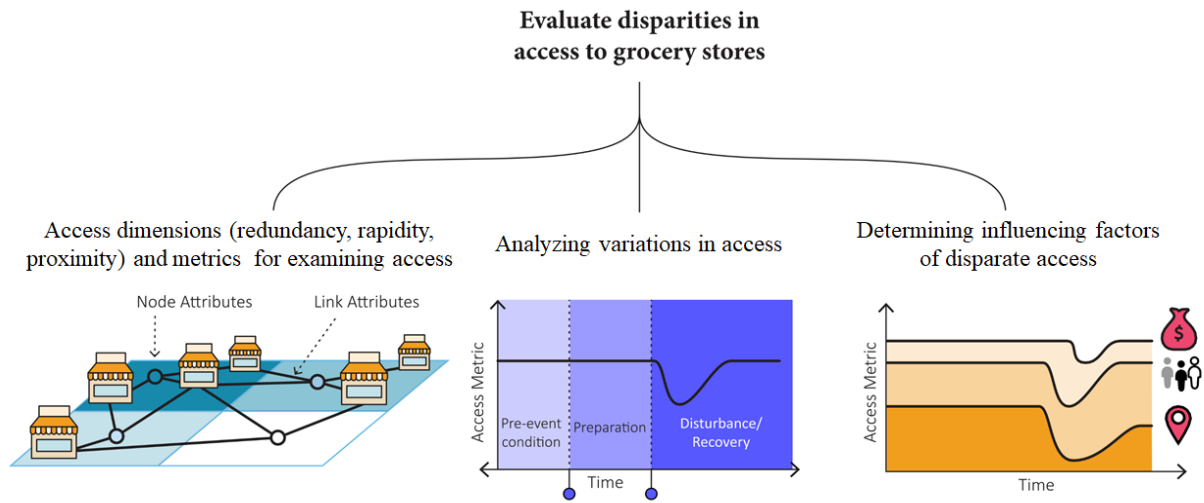

**Fig. 1 Framework for examining access disparities.** Three dimensions of access are examined based on network-based metrics. The fluctuations in access metrics are evaluated and compared across different sociodemographic groups to assess disparities.

*Study Context:*

This study examined the access to grocery stores in the context of Hurricane Harvey, which made landfall in Harris County, Texas, in August 2017. The storm dropped more than 60 inches of rain, triggering immense flooding and causing severe infrastructure disruptions[36,37]. Harris County area is prone to flooding impact and is frequently impacted by service disruptions caused by flooding events such as transportation disruptions and power outages. The direct flooding impact to the grocery stores and the loss of access due to disruptions in infrastructure systems such as road inundations and power outages caused significant disturbance to people's access to grocery stores. Furthermore, Harris County, within which the city of Houston is located, is a metropolitan area encompassing populations of a diverse range sociodemographic characteristics, providing a representative testbed for examining the equitable access to grocery stores in the face of natural hazards.

*Data:*

Data sources collected and analyzed in this study include location-based mobility data from Streetlight Data, points-of-interest (POI) visit data from SafeGraph, sociodemographic information from a 5-year estimate of the American Community Survey of the U.S. Census Bureau, Federal Emergency Management Agency (FEMA) Flooding Data, and the Food Access Research Atlas developed by United States Department of Agriculture. These data sources were aggregated at the census-tract level to construct the population-facility network models and to calculate access indicators and subsequent analyses. A detailed description of these data sources is provided below:

Location-based mobility data: The mobility data are provided by StreetLight Data, a commercial platform that provides origin-destination (O-D) analysis data. Our analysis aggregated anonymized data from cell phones and GPS devices to create travel metrics, such as duration and distance [38]. The O-D network of visits to grocery stores in Harris County was examined in this study from August 1 through September 30, 2017. Fig. 2 a shows the network of trips from traffic analysis zones (TAZs) to grocery stores during the second week of August 2017. These data incorporate the trips using different modes of transportation, including personal cars and public transit. By analyzing more than 40 billion anonymized location records across the United States in a month and enriching the analysis with other sources, such as digital road network and parcel data [39], StreetLight Data is capable of reaching approximately 23 penetration rate [40], covering distinct census divisions in North America's road network. Thus, the data provide a proper sample of human mobility in Harris County for examination of grocery store access disparity in the face of Hurricane Harvey.

Facility location data: POI data by SafeGraph were used to identify the location of facilities in this study. SafeGraph obtains location data by partnering with several location-based mobile applications. Data includes the basic information about the names, geographical coordinates, addresses, and North American Industry Classification System[41]. In this study, the top facility categories related to grocery stores, specialty food stores, and general merchandise stores (including warehouse clubs and supercenters, restaurants, and other eating places) were considered as the grocery stores in Harris County. Then we manually filtered the facilities to ensure that these POIs are stores used by the Harris County residents to obtain grocery needs. Fig. 2b shows the distribution of these POIs in Harris County.

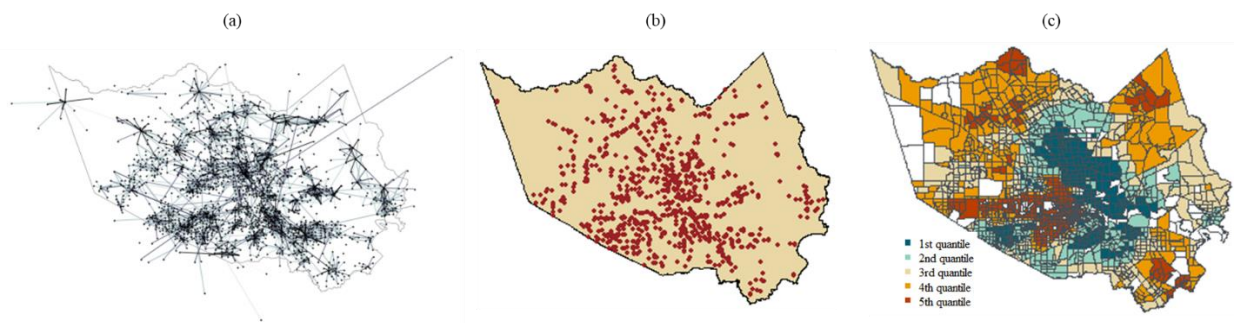

**Fig. 2 Distribution of the facilities and vulnerable population in Harris County: (a)** network of trips from 1804 traffic analysis zones to 920 grocery stores containing 6052 links (**a**); location of grocery stores in Harris County (**b**); distribution of income levels in Harris County (**c**); income is categorized into five quantiles, with the 5$^{th}$ quantile indicating the highest income level.

Sociodemographic characteristics: Sociodemographic characteristics of census tracts were determined by collecting data from the demographic characteristics estimate over the 2015–2019 period of the American Community Survey. The sociodemographic characteristics of the TAZs were determined from their overlap with census tracts. Fig 2c shows the income level of the TAZs using the census tract data from the 5-year estimate of the American Community Survey. Sociodemographic characteristics, such as socioeconomic status, household composition, minority status, and transportation from the census data, were used to examine grocery store access disparity among different sub-populations.

Flooding data: The Harvey Flood Depth Grid/Flood data from FEMA were adopted in this study to identify the flooded areas in examining the impacts of flooding on access [42]. The dataset was developed using the gage points from the National Weather Service and the terrain data from USGS. In this study, the flood data were examined to determine the flooding extent in each TAZ. Then those areas with a flooding extent greater than 6.5% of the area, which is the 75th-percentile flooding level in Harris County, were marked as flooded to further examine the relationship between flooding status and disrupted access to grocery stores.

Food access: The Food Access Research Atlas developed by USDA was used to identify areas labeled as having poor access to supermarkets in Harris County. According to the definition provided by the atlas [43], a census tract is considered to have poor access to grocery stores if at least 500 or 33% of residents live more than 1 mile (in urban areas) or 10 miles (in rural areas) from the closest supermarket, supercenter, or a large grocery store. This information was implemented to interpret the findings and to understand the extent that poor access to grocery stores contributes to the variation of access in the face of a disaster.

*Method*:

The analyzed data is an aggregation of the trip information from the users in home TAZs to the visited grocery stores. Based on the duration and time that a user spends in a particular location, Streetlight determines the home TAZ of the user. As an example of the aggregation, Streetlight first processes that there are 100 users residing in home TAZ A. Each user has a unique, one-way trip which is used to calculate the duration and distance between the home TAZ to the visited grocery store. An aggregated outcome could be that 40 users from home TAZ A attended Grocery Store A and 60 users from TAZ A attended Grocery Store B, where each user has a distinct duration and distance for their trip.

First, we constructed the population-facility network models of the study area. Population nodes are TAZs, facility nodes are individual POIs, and the links represent trips. Three separate network models were created: (1) one with unweighted links; (2) one with weighted links representing trip duration; and (3) one with weighted trip distance links. Accordingly, network-based metrics were used for examining the spatial and temporal patterns of access to grocery stores. Then, the variations in access to grocery stores during the different disaster phases were analyzed to characterize access to grocery stores, to examine access disparities, and to identify the factors contributing to such disparities.

Access indicators: In this study, we used network metrics for quantifying three distinct access indicators. These metrics, derived from the population-facility network models of visits to grocery stores, comprise topological and structural properties of the network of grocery visits. The three access metrics are: (1) the unweighted degree of TAZs (capturing number of unique POIs visited by the residents of a TAZ); (2) weighted degree of TAZs: trip duration (weighting indicates average trip duration; this indicator captures total average duration of residents at stores); and (3) weighted degree of TAZs: distance to stores (weighting indicates distance to stores; this indicator captures the total distance that residents of a TAZ take to access different stores). These metrics capture complementary characteristics of access, and are indicators of different access dimensions. The unweighted degree of TAZs captures *redundancy* for having access to grocery stores; the weighted degree based on duration captures *rapidity* of access and the weighted degree based on distance captures *proximity*. Table 1 summarizes the access indicators, dimensions of access, and the equations for measuring them based on the population-facility spatial networks. In these equations for calculating metrics for each TAZ ($i$), $a_{ij}$ are the elements of the adjacency matrix and $di\_w_{ij}$ and $du\_w_{ij}$ are the link weights based on the distance and duration of the trips from the TAZs, respectively, while $k_i$ represents the total number of trips.

**Table 1 Description of the developed access metrics in this study.**

| Dimension of Access | Access Indicator | Network Measure | Description | Equation |
|---|---|---|---|---|
| Redundancy | Number of POIs | Unweighted degree | Shows the number of unique visited grocery stores. | $\sum_j a_{ij}$ (1) |
| Rapidity | Duration (sec) | Average weighted degree of duration | Measures the average trip duration in seconds. | $\frac{\sum_j du\_w_{ij}}{k_i}, k_i = \sum_j a_{ij}$ (2) |
| Proximity | Distance (mile) | Average weighted degree of distance | Measures the average trip distance in miles. | $\frac{\sum_j di\_w_{ij}}{k_i}, k_i = \sum_j a_{ij}$ (3) |

The access indicators (Table 2) capture properties of access in different phases of the disaster. The number of unique visits to POIs informs about available options, which could provide *redundancy* in the face of disasters. Other access indicators (duration and distance), which capture the *rapidity* and *proximity* access dimensions, are influenced by human activities (lifestyle patterns) and infrastructure conditions (facility locations and road congestion). Table 2 summarizes the interpretation of these access indicators during the normal period, preparation, and impact/short-term recovery period. The normal period establishes the baseline access indicators. However, the preparation and impact/short-term recovery interpretations focus on the variation in access compared to the baseline (normal) period. This comparison with the baseline period provides insights regarding the effects of disturbance on residents' access to stores during the preparation and impact/short-term recovery period.

**Table 2 Interpretation of access indicators during different disaster phases.**

| Dimensions of Access | Access Indicator | Normal period | Preparation | Impact/Short-term Recovery |
|---|---|---|---|---|
| Redundancy | Number of POIs | More visited stores means more options to satisfy life needs | Increased redundancy means more visits compared to baseline due to higher preparation levels. | Increased redundancy compared to baseline means greater recovery activity. |
| Rapidity | Duration (sec) | Greater duration means a lower rapidity and longer commutes to grocery stores due to longer | Reduced rapidity means a greater duration compared to baseline due to more time spent obtaining supplies as a result of | A lower rapidity (greater duration) compared to baseline means greater impact to access (due to store and/or road inundation). |

| | | distance or traffic congestion. | longer commutes and traffic congestion. | |
|---|---|---|---|---|
| Proximity | Distance (mile) | Greater distance means a lower proximity and thus greater distance from stores. | Closer proximity means greater distance compared to baseline, which means greater effort for obtaining supplies. | A lower proximity (greater distance) compared to baseline means greater impact to access (due to store and/or road inundation). |

*Examining spatial and temporal patterns of variation in access*: The examination of access to grocery stores based on the adopted access indicators was conducted across three disaster phases: prior to the event (without a disturbance to the network), during the period from the formation of Hurricane Harvey to prior to landfall (focus on preparedness activities), and during the landfall and recovery. The spatial and temporal patterns of impact and short-term recovery are evaluated to discover disparate access during different disaster phases. Then unsupervised learning approaches for time-series clustering characterize patterns of impact and short-term recovery on access indicators. Then, we performed statistical analysis to evaluate the influence of different factors (i.e., sociodemographic characteristics, location in a food desert, and flood status) on residents' access to grocery stores at different stages of disaster. The analysis steps are explained in the remainder of this section.

Constructing the population-facility spatial network: First, we constructed the population-facility network of commutes from TAZs ($i$) to grocery stores ($j$). This TAZ-POI network is a directed and bipartite network mapped based on the geographic coordinates [44]. The constructed network represents trips from TAZs to POIs as the links with weights ($w_{ij}$) based on distance and duration metrics. The three access indicators (i.e., visited POIs, duration, and distance) are determined based on a daily aggregated trip numbers on each link between TAZ and POI pairs.

*Calculating the percentage change of access indicators:* The percentage change of each access indicator is calculated based on comparing the daily values with a defined baseline. The baseline period for each indicator at a TAZ is calculated considering each day as a unit, as a weekly pattern was observed in the data. The baseline period (August 1 through August 20, 2017) includes three weeks to define the weekdays' baseline values. Then the percentage change for each TAZ was calculated based on the following equation:

$$Pc_{i,d} = \frac{M_{i,d} - B_{i,d}}{B_{i,d}} \qquad (4)$$

where, $Pc_{i,d}$ is the percentage change in the access indicator at TAZ ($i$) in a date ($d$). $M_{i,d}$ is the access indicator, and $B_{i,d}$ is the calculated baseline value for weekday corresponding to the date for determining percentage change. Then the resulting time series is used for examining the variations in access to the grocery store across different phases.

*Dealing with missing data:* To deal with missing data on certain dates for the three indicators, we first filtered those TAZs which did not have data for 5 consecutive days. Then, the Kalman imputation method was applied to the data to deal with the missing data in the time series. The ImputeTS package in R was used to implement the algorithm on the univariate time-series for all the access indicators[45]. The Kalman filter method uses the structural time series ideas where the system is outlined by a well-defined model with unknown parameters [46]. The maximum likelihood approach was implemented to determine the time-dependent model parameters [47].

*Time-series clustering of TAZs based on their access indicators:* We implemented time-series clustering algorithms on the access indicators to characterize and examine the spatial and temporal variations. First, a 3-day moving average was applied to the time-series data of each access indicator to extract trend and limit noise. Then, a partitional clustering using dynamic time warping (DTW) distance in dtwclust in R [48] was used to perform the time-series clustering. Partitioned procedures were considered to be optimization problems that maximize the inter-cluster distance and minimize the intra-cluster distance[49]. DTW is a widely used approach for defining the distance in time-series data [50]. In this study, we implemented a multivariable time-series clustering approach for classifying the TAZs. The 3-day moving average time series related to the access indicators for each TAZ were used to identify the TAZ clusters. Then, two cluster validity indices, COP and modified Davies-Bouldin, were used to determine the number of clusters [51].

**Results:**
The calculated access indicators for the TAZs (described in the Method section) were used to examine the access at three disaster phases: pre-disaster condition, during preparation, and during disturbance and recovery of the affected areas. The analysis covers a period between August 1 and September 15, 2017. The pre-disaster period (August 5 through August 16, 2017) covers the period immediately prior to landfall; however, no disturbance had occurred to the residents' access to grocery stores. This period captures access to grocery stores during normal conditions. Following the formation of Hurricane Harvey and the issuance of the hurricane watch on August 23, preparation activities were initiated, and some grocery stores were out of stock due to increased demand. Finally, Hurricane Harvey made landfall in Harris County on August 25, 2017, causing several road inundations and disrupting access to grocery stores, which affected Harris County residents' access.

*Pre-disaster normal period*: To understand the key characteristics of the access in the normal condition, we analyzed the period before the Harvey landfall. Dynamic clustering was used to identify the distinct clusters of TAZs with similar access characteristics for each indicator. The results show the patterns of access to grocery stores for the different TAZs in Harris County. Fig. 3 depicts the identified clusters for the three access indicators together with the maps of these clusters (darker colors schemes show a greater value). These plots also show a spatial clustering pattern which suggests that areas in the proximate TAZs have similar access characteristics. The number of visited stores indicator, which captures the *Redundancy* dimension of access, shows a spatial clustering pattern in some areas. The TAZs in the East and South of Harris County show a lower number of visited stores, which translates to a low *redundancy* of access to grocery stores. The difference in the *redundancy* in access to grocery stores could be related to the number of nearby stores, the size of the stores, and households' lifestyles and shopping behaviors.

On the one hand, the more stores available in the proximity of a TAZ, the more the visits would take place during normal conditions. On the other hand, the size of stores could also affect the frequency of visits to stores, as larger grocery stores with a larger, more varied inventory could meet diverse needs of the households; one visit would be adequate to satisfy the weekly needs for groceries and other supplies. Furthermore, some households rely on coupons or have lifestyles that require them to follow certain shopping behaviors that affect the number and type of grocery stores they visit. Therefore, the properties of the facilities, their distribution, and the lifestyle characteristics of the households could influence the access *redundancy* indicator. *Proximity* and *rapidity* indicators also show the presence of some spatial clusters, which suggests TAZs would have similar access indicators to their neighboring TAZs. Examining the three access indicators together, however, shows little overlap among the clusters for the three indicators. This result shows the spatial heterogeneity of TAZs in terms of their three access indicators. Fig 3d shows the aggregated map and clusters of TAZs based on the three indicators. In this map, the high and low clusters of the access indicators are used for characterization of access during normal times. The most frequently occurring categories are those with a low number of visits, low duration, and a low distance followed by those areas with a high number of visits and low duration and distance. Both of these clusters indicate good access. These categories (highlighted by light blue and green) with proper *rapidity* and *proximity* access are more concentrated in high-income areas west and southwest of downtown. The large TAZs in the north (highlighted in red) are the next most frequently occurring clusters with high levels of duration and distance and a low level of number of visits, which shows poor access in terms of *proximity*, *rapidity*, and *redundancy*.

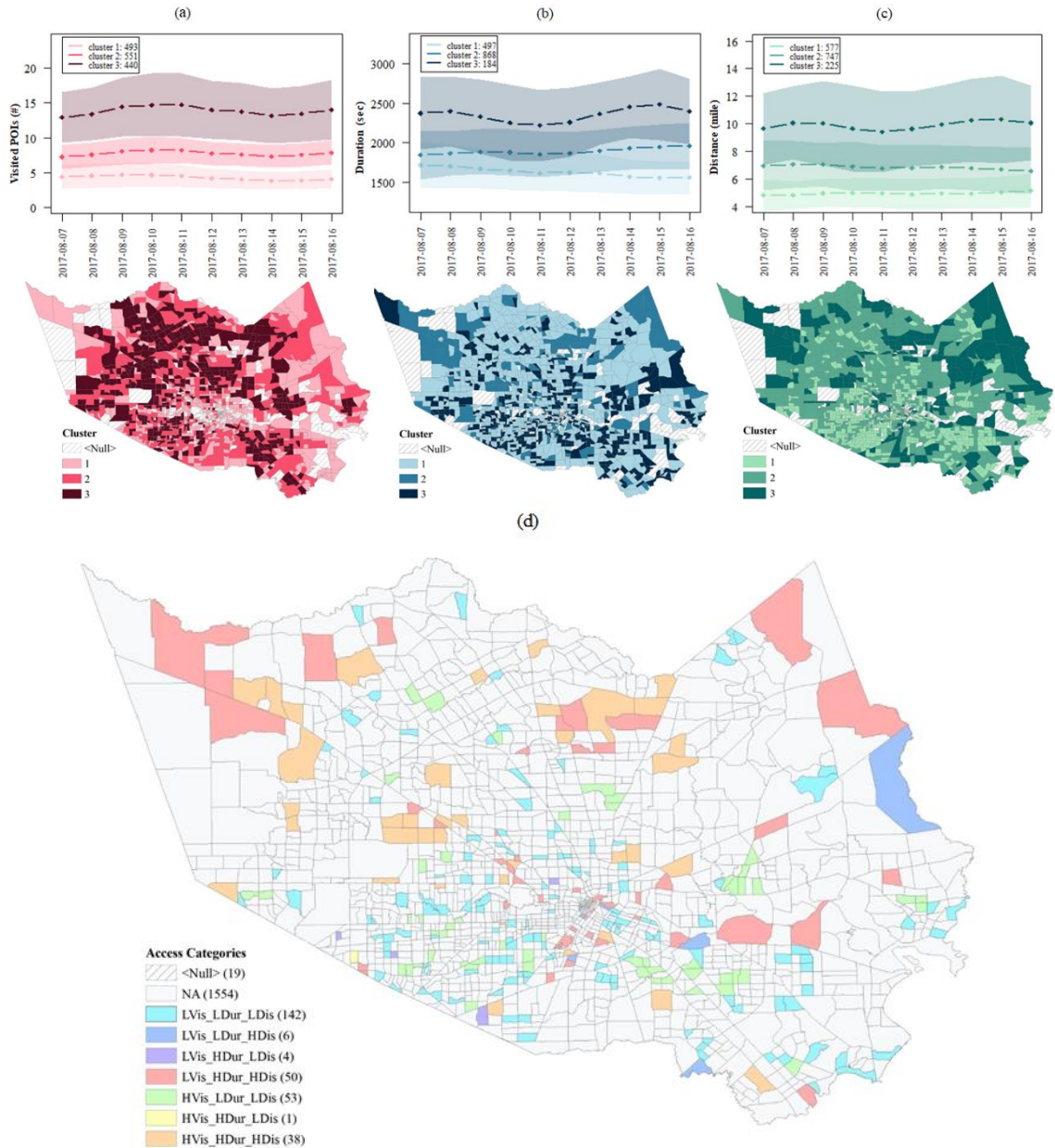

**Fig. 3 Clusters of access indicators during the pre-disaster normal condition.** These figures show the normal condition (August 7–16) daily mean values of the indicator for each cluster in addition to the *1* standard deviation uncertainty band. The clusters show the distinct temporal and spatial patterns for access indicators: (**a**) through (**c**) values for visited POIs, duration, and distance respectively; areas with a greater number of visited stores have a greater *redundancy*, while those areas with a greater level of duration and distance have a lower level of *rapidity* and *proximity* consecutively; categorized areas based on the access indicators of the high (H) and low (L) levels; the number of TAZs in each group are shown in parenthesis (**d**).

The association between the sociodemographic characteristics of the TAZs, as well as the facility distribution characteristics (i.e., number of stores in the TAZ and location in a food desert) with the access indicators, were examined to evaluate factors influencing spatial variations of access patterns. The results show that the identified clusters have distinguishable demographic and facility distribution characteristics. Table 3 shows the results of comparing the sociodemographic and facility distribution characteristics related to all access indicators for
11

different clusters. ANOVA and chi-square tests were implemented to test if differences between the clusters are statistically significant at a 0.05 confidence level.

The results for the number of visited POIs indicator show that the cluster 1, which the highest number of unique stores visited (highest redundancy), had a lower socioeconomic status. These groups of people have diminished access to big supermarkets and are more likely to buy their needed supplies by using coupons from different stores to save money. In addition, cluster 1 has more stores in the TAZs and also a lower chance of being located in a food desert (being far from a store) compared to other clusters. The availability of stores also partly explains a greater number of unique grocery stores visited by residents of TAZs in this cluster. The results related to the *proximity* and *rapidity* access dimensions do not show a clear disparity with respect to the sociodemographic characteristics. In fact, the results show that TAZs with a better socioeconomic status have lower proximity and rapidity access dimensions. This pattern could also be due to the fact that these TAZs are more likely located in residential areas, and their residents have a greater capability to commute longer distances to obtain their needed supplies from specific stores. In addition, these clusters are located in areas with a lower number of stores and a higher chance of being located in a food desert.

**Table 3 Properties of the clusters of access indicators during the normal period.**

| Variables | Unique POIs visited (redundancy dimension) | | | Duration (rapidity dimension) | | | Distance (proximity dimension) | | |
|---|---|---|---|---|---|---|---|---|---|
| | C_1 | C_2 | C_3 | C_1 | C_2 | C_3 | C_1 | C_2 | C_3 |
| *Sociodemographic characteristics* | | | | | | | | | |
| Per capita income | 36008.19* | 28305.49 | 28305.94 | 33211.45 | 30418.46 | 35022.94* | 31751.14 | 31943.62 | 31850.78 |
| % people below poverty line | 17.84 | 18.76 | 18.29 | 18.35 | 18.46 | 16.73 | 19.67* | 17.46 | 17.02 |
| % people without diploma | 20.06 | 21.41 | 22.35 | 21.07 | 21.39 | 18.84 | 22.83* | 20.38 | 18.24 |
| % people over 65 | 10.58* | 9.77 | 8.81 | 10.32* | 9.42 | 8.84 | 10.04* | 9.56 | 8.86 |
| % people with disability | 10.31* | 9.77 | 8.81 | 10.00* | 9.61 | 9.13 | 9.83 | 9.60 | 9.52 |
| % people with minority | 64.90 | 70.44 | 71.03* | 66.21 | 69.42* | 66.86 | 69.09 | 67.84 | 66.31 |
| % housing units crowded | 5.53 | 6.20 | 6.70* | 5.92 | 6.21 | 5.38 | 6.75* | 5.71 | 5.13 |
| % households without vehicle | 7.36 | 6.79 | 6.39 | 7.29 | 6.88 | 6.38 | 7.89* | 6.59 | 5.83 |
| *Facility distribution characteristics* | | | | | | | | | |
| *Location in a food desert* | *0.45** | *0.38* | *0.34* | *0.32* | *0.40* | *0.52** | *0.23* | *0.43* | *0.62** |
| *Number of stores* | *0.67* | *1.03* | *1.87** | *1.11* | *1.23** | *0.81* | *1.17** | *1.24* | *0.73* |

Note: * shows the highest value for the comparisons, which are significant at a 0.05 confidence level through the ANOVA or chi-square test.

*Preparation:* The percentage change in access indicators in comparison with the defined baselines was examined to assess the disparities in access to grocery stores in the preparation phase. In the time span between the identification of Hurricane Harvey until landfall in Harris County, residents attempted to store supplies, such as food and bottled water, to protect their households against hardship. Since there was limited mandatory evacuation by public officials, most residents sheltered in place; thus, storage of food and bottled water was critical to riding out the storm. Many grocery supplies were out of stock due to the high demand and the shelter-in-place protective behavior of the residents, which disturbed the access patterns to grocery stores. We analyzed the percentage change in the access indicators to seek out disparities in access during the preparation phase. The percentage change of the access indicators on the day before the hurricane landfall (August 24, 2017) was chosen to examine disparities in access patterns. The examination of the daily patterns of access indicators showed that their greatest percentage change occurred on this date, and residents showed a high level of activity to obtain grocery supplies for their households.

To explain the variations in the access indicators during the preparation period, we first examined the patterns in the pre-disaster level for the three indicators (as discussed in the previous section) together with income level. The examination of the association of these two factors with the percentage change in the indicators reveals that there is a statistically significant relationship between the access patterns in the normal condition with the percentage change of these indicators during preparation stage. Results show that those with lower accessibility based on the three indicators, meaning fewer unique POIs visited, longer duration, and longer distance, had a higher percentage change during disaster when compared to normal conditions. This means that those who visited few number of POIs (redundancy) greatly increased the number of unique visits. A similar pattern exists in the *rapidity* dimension, as the TAZs with a longer trip duration (lower *rapidity*) in the normal condition showed a lower increase in the percentage change. The increase in the trip distance also shows that those TAZs which have shorter distance (better *proximity*) in the normal condition have a greater increase in the preparation period.

Within the low, median, and high levels of accessibility in the normal condition, the low income (bottom 33%) and high income (top 33%) TAZs were compared to their new levels of accessibility in the preparation stage. The results show disparate access to grocery stores during the preparation phase of Hurricane Harvey. Fig. 4 shows the boxplots of the percentage changes in the access indicators. ANOVA and Tukey's tests show that there is a significant difference in the percentage change of the access indicators across income groups at 0.05 confidence level. Regarding redundancy indicator, results (Fig. 5) show that the high-income group had a significantly higher increase in visits to unique stores compared to lower-income groups. This result indicates that high-income groups were able to improve their redundancy by visiting more unique stores during the preparedness stage to supply their households with adequate groceries before the hurricane made landfall in Harris County. Regarding the rapidity indicator, lower-income groups who already had long duration trips in the normal condition had disproportionate increase to their trip durations during the preparation period compared to high income TAZs. Regarding the proximity indicator, high-income groups who initially had a low distance to grocery stores during the normal condition had the greatest increase in their distance during the preparation stage. This shows the capability of high-income groups to commute further to obtain their supplies. However, the increase in the percentage change of the distance is not significantly different across the income groups for TAZs with low *proximity* (long distance) in the normal condition.

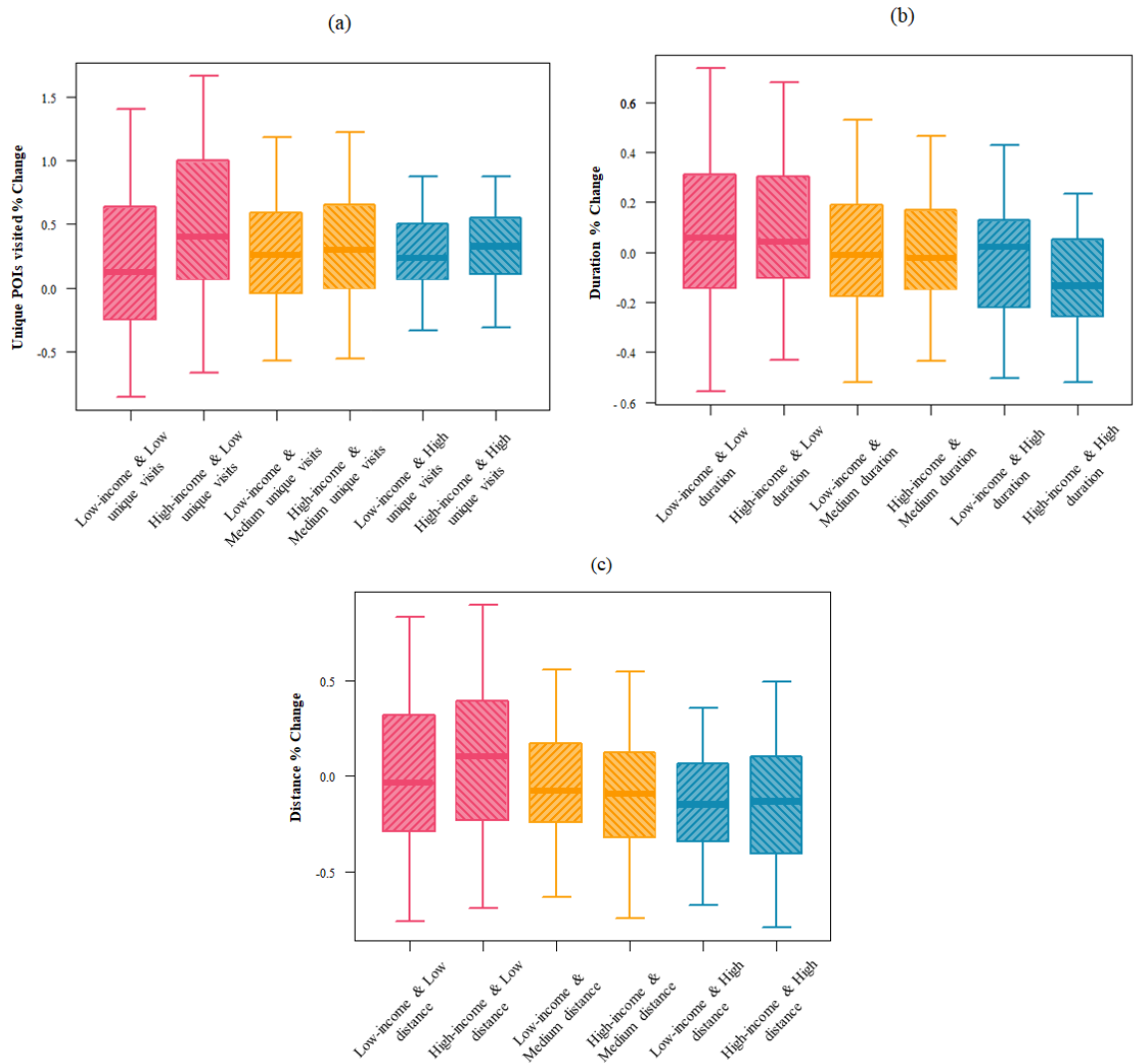

**Fig. 4 Comparison of the percentage change of the access indicators**. These boxplots show the percentage change of the access indicators for the day before the hurricane landfall. The six groups categorized based on the income and pre-disaster access level show disparate access variations during the preparation phase: percentage change in the number of unique visited stores (**a**); the redundancy dimension of access also increase as this value increases; percentage change in the distance which has a negative relationship with the rapidity dimensions of access(**b**); percentage change of distance (**c**), the proximity access dimension decreases as the distance increases.

In the next step, we examined the effect of facility distribution factors, namely the number of POIs in a TAZ, on the access indicators. First, the correlation between the numbers of POIs in a TAZ with the access indicators (Table 4) shows that there is only a slight but significant association between the percentage increase in redundancy, or number of unique visits, and the number of POIs. TAZs with more POIs do not show a large increase in visits as they already provide more options for the residents. However, the associations between the number of POIs and the rapidity and proximity metrics are not significant. The results indicate that change in access indicators during preparedness stage is not associated with the number of POIs. This result shows that, during preparation stage and due to surge in demand for grocery supplies, residents need to take longer and further commutes to POIs outside their TAZs to obtain supplies regardless of the number of grocery stores in their own *proximity*. Thus, those residents (low-income groups, as shown in earlier results) who could not increase their access distance may not be able to adequately prepare for the impending hazard.

**Table 4 Correlation analysis for the association between the number of POIs in a TAZ and access indicators.**

|  | Visited POIs (% change) | | Duration (% change) | | Distance (% change) | |
| --- | --- | --- | --- | --- | --- | --- |
|  | *Coefficient* | *P-value* | *Coefficient* | *P-value* | *Coefficient* | *P-value* |
| Number of POIs in a TAZ | -0.06 | 0.007 | -0.01 | 0.448 | -0.03 | 0.100 |

In the next step, we categorized TAZs into four categories based on income in conjunction with location in a food desert to understand the effect of these factors on access indicators. Fig. 5 shows boxplots of the percentage change in the access indicators; the ANOVA test suggests a significant difference in access across the four categories. The results show an interaction between income and food deserts with access to grocery stores. High-income TAZs located in a food desert show a higher number of visited POIs, while the duration and distance are not significantly higher than that of low-income TAZs in a food desert. Within food deserts, these high-income TAZs were able to visit more stores to gain their grocery store needs without having to increase their trip distance and duration. Thus, being located in a food desert, while affecting people's access to the grocery stores to some extent, does not have an equal impact on the access of different income groups. Meanwhile, those high-income TAZs not in food deserts were also able to access farther POIs which means high income were able to access grocery in and outside their close proximity. Integrating the results of food desert status with the sociodemographic information, the access inequalities further reveal themselves. Also, this result shows that being located in a food desert is not adequate to evaluate residents' access to grocery stores during the preparation stage of disasters; access to grocery stores during preparedness stage is more influenced by the dynamics of human protective actions influenced by capabilities.

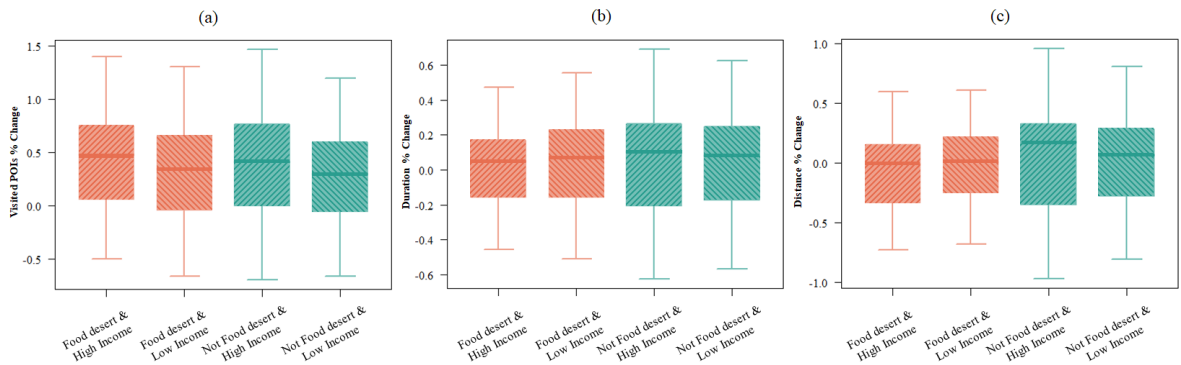

**Fig. 5 Comparison of the percentage change of the access indicators**. These boxplots show the percentage change of the access indicators for the day before the hurricane landfall. The four groups categorized based on food desert status and income show those TAZs with a high income and not located in a food desert have the greatest change in their access indicators during preparedness stage: values for the number of unique visited stores (greater number of visited stores means more access redundancy) (**a**); values for the duration of trips (greater trip duration means less rapidity) (**b**); values for the distance (greater distance means less proximity) (**c**).

*Impact/short-term recovery:* To understand the dynamic spatiotemporal patterns of impact and short-term recovery of access to grocery stores, we conducted multivariable temporal clustering on the three access indicators. The results indicate the presence of two clusters of access patterns to grocery stores in TAZs whose access was affected by flooding. Fig. 6 shows the dynamic pattern of access indicators for the identified clusters. TAZs in cluster 2, which are depicted in blue in Fig 6a, have better access and show a pattern that has a greater increase in the visited POIs indicator (increase in *redundancy*); however, the cluster shows a lesser increase for the distance (decrease in *proximity*) and duration (decrease in *rapidity*) indicators during the recovery period. In particular, the *proximity* indicator seems to show a lesser level of decrease (increase in distance) in cluster 2 compared to cluster 1. The patterns suggest that while cluster 2 visited more stores, their duration and distance did not increase significantly compared to cluster 1. This result shows that cluster 2 TAZs had better access, as they could meet their needs by visiting more stores without the need to significantly increase their trip distance. Increased *redundancy* without the need for decreasing *proximity* is an indication of better access for cluster 2 (compared with cluster 1, which did not have significant increase in access *redundancy* but had significant decrease in access *proximity*).

Fig 6d shows the maps of the identified clusters; this map suggests that the TAZs in clusters form spatial cliques. These cliques show a distinct pattern while being in proximity of the cliques of other access clusters. To further examine factors affecting access patterns in each cluster, we examined the sociodemographic characteristics and the facility distribution characteristics, such as number of stores and location in a food desert.

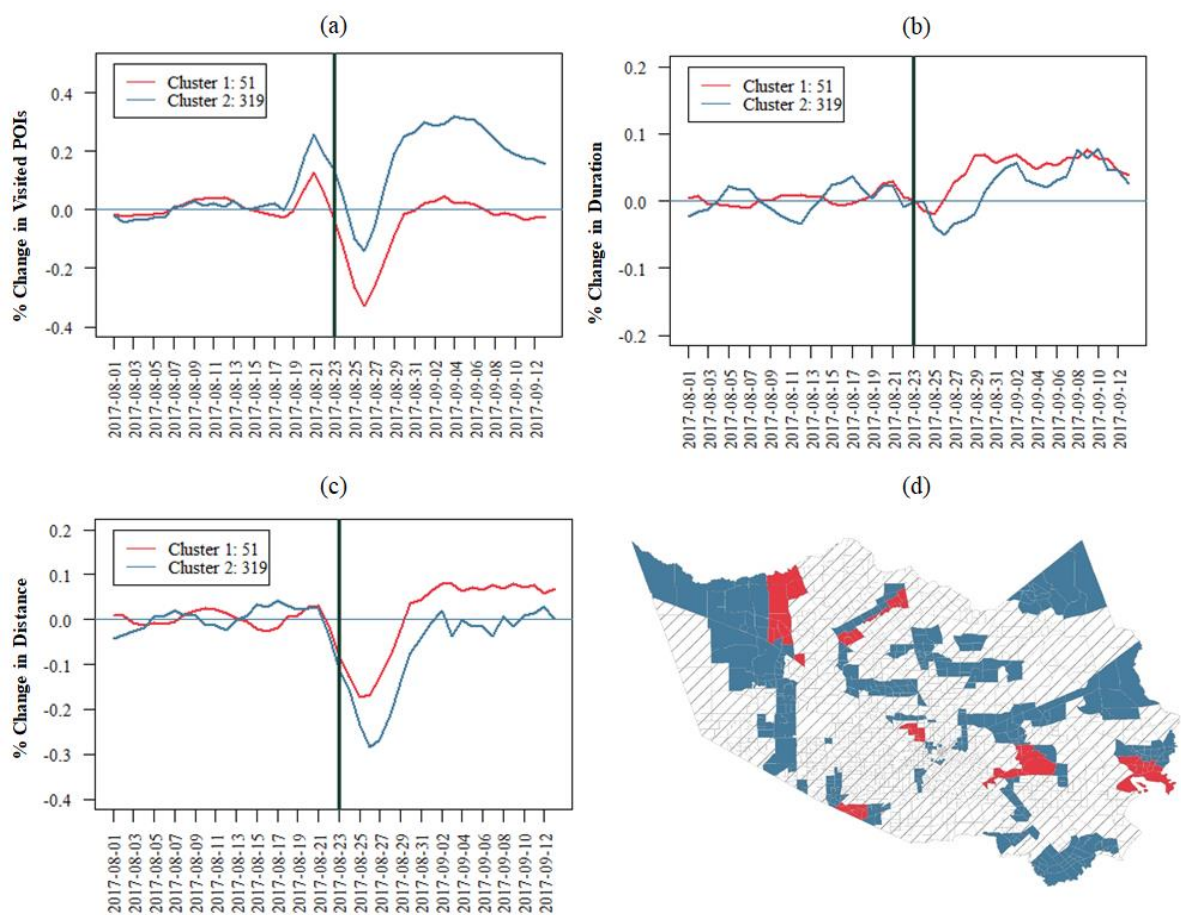

**Fig. 5 Clusters of access to grocery stores by considering the three access indicators**. These clusters include areas affected by flooding: (**a**)–(**c**) time-series level of the identified clusters for the access indicators; (**d**) maps of the clusters; (**a**) daily pattern of impact and short-term recovery for the visited POIs, which shows the patterns of the change in the redundancy access dimension compared with the baseline. In (**b**) and (**c**) show that as duration and distance increase, rapidity and proximity decrease. The cluster identified as blue shows an increased level in visited POIs, while the distance and duration for this cluster show a better access pattern compared with the red cluster.

Fig. 7 shows the comparison of the sociodemographic characteristics and the facility distribution in the two clusters. These boxplots (Fig 7 a–c) are the results of the comparisons among the TAZs in each cluster (significant at 0.05 confidence). The results suggest that the TAZs in cluster 2 with better access have higher income and a lower percentage of the minorities. This result indicates that in the aftermath of disasters, the socially vulnerable population (low-income and minority groups) has lesser access to grocery stores. In addition, those TAZs in cluster 2 had a higher chance of being in a food desert and have a lower number of POIs in their TAZ, they have better access in terms of improved access redundancy without decreased access proximity.

The distribution of income level within a food desert in the two identified clusters was examined in a proportion test analysis. The results of the proportion test show that in cluster 2, the proportion of high-income TAZs in food deserts is significantly higher than that of low-income TAZs in food deserts with proportions being 0.53 and 0.25, respectively (p-value of less than 0.001). However, such a distribution did not exist in cluster 1, with proportions being 0.57 and 0.47 for the high-income and low-income TAZs, respectively (p-value of 0.76). Thus, the high proportion of higher-income areas in food deserts allowed for better access in cluster 1 than in cluster 2 despite a higher level of being in a food desert. These results suggest that the food desert designation is not adequate to capture residents' access to grocery stores in the aftermath of disasters, and more data-driven insights (such as the results of this study) are needed to evaluate the access of different sociodemographic groups to critical facilities in times of disaster. Furthermore, the results from the comparison of the extent of flooding in the TAZs do not show a significant difference between the two clusters (p-value = 0.33). The insignificant difference in flooding shows that the diminishment of access to grocery stores goes beyond the extent of flooding; however, the sociodemographic characteristics and the facility distribution characteristics explain the disparate access to grocery stores for the two groups.

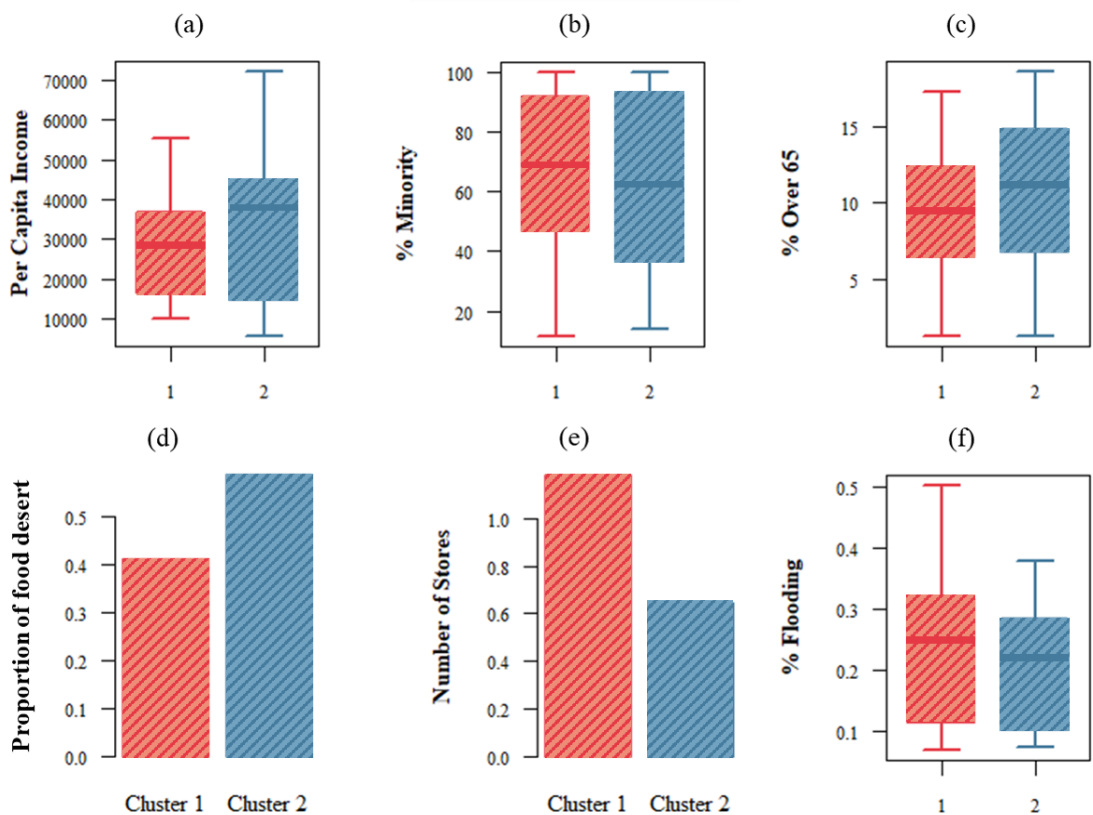

**Fig. 7 Properties of the identified clusters based on the multivariable dynamic clustering.** Figures **a-c** show the unequal level of income, percentage of minorities, and percentage of elderly in the two clusters: the proportion of being in a food desert is higher in cluster 2 (**d**), and there are more grocery stores in the red cluster (**e**). Lastly, (**f**) shows that the level of flooding is not statistically different among the two groups.

**Discussion and Concluding Remarks:**
In this study, we analyzed high-resolution location-based data to characterize and uncover the disparities in access to grocery stores before, during, and after disaster events. We developed and implemented access indicators based on the population-facility network structure and analyzed their spatial and dynamic patterns to characterize access to grocery stores and reveal the presence of disparities at different stages of a disaster. The high-resolution location-based data enabled capturing the patterns of trips of residents to grocery stores and facilitated the characterization of access based on observational data. The current approaches in examining access based on physical and distance-based metrics fail to provide adequate and reliable information about the properties of access during a disaster to enable decision-making for equitable access.

The key findings of the research show the importance of characterizing access based on the three dimensions of *redundancy*, *proximity*, and *rapidity* since there are unique differences to the normal conditions to the disaster setting. During the normal time, areas of lower socioeconomic status have a comparatively higher *redundancy* and slightly lower levels of *proximity* and *rapidity* compared those with higher socioeconomic status. A possible explanation is that lower socio-economic groups tend to have lesser access to the large supermarkets that provide a large variety groceries and other supplies, making them more dependent on visiting multiple stores. This sub-population is more dependent upon public transportation and less capable of making longer commutes. In addition, a combined examined of dynamic patterns of access indicators revealed that the disaster disproportionately exacerbated access disruptions for socially vulnerable groups, in the context of the 2017 Hurricane Harvey in Harris County, Texas. In the preparation period prior to the disaster, areas of higher income had a greater increase to the number of unique visits (*redundancy*) and the distance traveled (*proximity*). This reveals the higher capability to take protective actions and seek essential resources for preparation in comparison to lower-income areas. This phenomenon changes in the impact/ short recovery stages of the disaster setting. Areas with higher income and a lower percentage of racial minorities maintained their higher redundancy, meaning they were able to access more stores similar to the preparation periods. In this case, though, these areas had a lesser decrease in distance (*proximity*) and less trip duration (*rapidity*) compared to those of lower income and larger minority population. Thus, these groups could visit more stores without longer and more lengthy commutes.

The examination of different dimensions of access in the context of disasters confirmed the inadequacy of physical measures of access, such as location in a food desert and the number of available stores in an area, for understanding access in disasters. The research findings show that while these factors are better capable of explaining the variations in access during normal times, they fail to explain access characteristics during different stages of a disaster. The multivariate analysis of access showed that the areas with higher income and a lower percentage of racial minorities have better access to grocery stores. However, higher income areas tended to have a lower number of stores in their area and a higher likelihood of being in food deserts. This could be because higher income areas live in residential areas, where they are more separated from grocery stores; however, they have the capability to travel further distances and visit more stores. These findings indicate that, while focusing on facility distribution characteristics is useful in understanding the disparities during the normal condition and for purposes of urban development, the standard physical measures of access, such as location in a food desert and the number of available stores in an area, is not sufficient for understanding access to facilities in the context of disasters.

Although the research captured an element of disparity in the accessibility to grocery stores, it is important to acknowledge the complex metrics that cumulate in food inaccessibility at the individual household level. Thus, future research could integrate the significant findings into other physical and social indicators. For example, financial ability, particularly the capabilities or purchasing power of households to understand the challenges and decision-making of individuals. Data-driven evaluation of access and disparities in disasters could inform future emergency response, preparedness, and mitigation plans and actions. For example, the placement of temporary distribution centers in areas with lowest levels of access could help with better and more equitable preparedness for shelter-in-place populations. In using the current case study, it was established that Cluster 1 had less accessibility metrics when compared to Cluster 2. Temporary distribution centers could be prioritized in the highlighted areas of cluster 1 (Figure 6). Additionally, infrastructure development and improvement plans could be informed by the findings of this study for equitable allocation of resources to enhance access to grocery stores. For example, roads which are critical for providing access to facilities can be identified and retrofits/enhancements could be implemented to mitigate flooding of those roads. The development of future facilities can be informed by the results of this study to enhance access in a more equitable manner both in normal times, as well as during disasters.

**Data availability:**
The data that support the findings of this study are available from SafeGraph and Streetlight Data, but restrictions apply to the availability of these data, which were used under license for the current study. The data can be accessed upon request submitted on SafeGraph and Streetlight Data. Other data we use in this study are all publicly available.

**Code availability:**
The code that supports the findings of this study is available from the corresponding author upon request.

**Acknowledgment:**
The authors would like to acknowledge the funding support from the National Science Foundation CAREER Award under grant number 1846069. The authors would also like to acknowledge SafeGraph and Streetlight Data for providing the points-of-interest and traffic data. Any opinions, findings, conclusions, or recommendations expressed in this research are those of the authors and do not necessarily reflect the view of the funding agencies.

**Competing Interests:**
The authors have no competing interests to address.

**Author Contributions:**

All authors critically reviewed and revised the manuscript, gave final approval for publication, and agree to be held accountable for the work performed therein. A.E was the lead researcher and first author who was responsible for the location-based mobility data analysis, interpretation of results, and manuscript composition. N.C provided further support in the clarification of the research scope, methods, and results. F.Y contributed to the research design and manuscript. X.X assisted in the data processing and data analysis. A.M was the faculty advisor of the project and was significant in framing the research and manuscript writing.